\documentclass[aps,prd,twocolumn,nofootinbib,groupedaddress,amsfonts,floatfix]{revtex4}

\usepackage{graphicx,amsmath,amssymb,amstext}
\usepackage{amssymb,amsbsy,amsfonts,amsthm,hyperref}
\usepackage{lscape,placeins}

\usepackage{color}
\usepackage{ifthen}
\newboolean{editorial}
\setboolean{editorial}{true}
\newcommand{\editorial}[2]{\ifthenelse{\boolean{editorial}}{\textcolor{red}{[\textsf{\textbf{{#1}}}: }\textcolor{blue}{\textsf{{#2}}}\textcolor{red}{]}}{}}

\begin{document}

\title{Gravitional radiation from first-order phase transitions in the presence of a fluid}

\author{John T. Giblin, Jr${}^{1,2}$}
\author{James B. Mertens${}^2$}

\affiliation{${}^1$Department of Physics, Kenyon College, 201 N College Rd, Gambier, OH 43022}
\affiliation{${}^2$CERCA/ISO, Department of Physics, Case Western Reserve University, 10900 Euclid Avenue, Cleveland, OH 44106}

\begin{abstract}
First-order phase transitions are a source of stochastic gravitational radiation.  Precision calculations of the gravitational waves emitted during these processes, sourced by both the degrees of freedom undergoing the transition and anisotropic stress of the coupled, ambient constituents, have reached an age of maturity.  Here we present high-resolution numerical simulations of a scalar field coupled to a fluid and parameterize the final gravitational wave spectrum as a function of the ratio of the energies of the two sectors and the coupling between the two sectors for a set of models that represent different types of first-order phase transitions.  In most cases, the field sector is the dominant source of gravitational radiation, but it is possible in certain scenarios for the fluid to have the most important contribution.
\end{abstract}

\maketitle

\section{Introduction}
\label{intro}

As we prepare for the first direct detection of gravitational radiation, there is increasing interest in analytic and numerical models of gravitational wave sources.  These sources generally fall into two categories: transient events that can be identified by waveforms, and stochastic backgrounds.  Many cosmological processes fit into this latter category, including phase transitions, preheating, and cosmic string networks -- just some examples of processes that produce potentially observable gravitational wave signatures.

For a cosmological process that generated gravitational waves during the early radiation-dominated phase of the Universe, it is now well known that we can estimate the current-day peak frequency $f_0$ of the gravitational waves by estimating the peak wavenumber $k_*$ of the sub-horizon modes at the time the waves were generated.  Taking these modes to be a multiple of the Hubble scale at the time of the process, $k_* = \zeta H_*$ where $\zeta \sim 1-100$, the transfer function \cite{Easther:2006gt}
\begin{equation}
\label{freqestimator}
f_0 = \frac{6\times 10^{10}k_*}{\sqrt{m_{\rm pl} H_*}}\,{\rm Hz} = \left(\frac{\rho_*^{1/4}}{m_{\rm pl}}\right)\zeta\,1.7\times10^{11}\,{\rm Hz}
\end{equation}
determines the ideal observational frequency with only an estimate of $\zeta$ and knowledge of the energy scale of the Universe at the time of the process, $\rho_*$.

However, determining the present day amplitude and the precise shape of the spectrum is a complex process.  In recent years, numerical techniques have promoted analytic calculations and semi-analytic estimates to precision predictions.  Processes with a single scale, e.g. preheating \cite{Traschen:1990sw,Kofman:1994rk,GarciaBellido:1997wm,Khlebnikov:1997di,Greene:1997ge,Parry:1998pn,Bassett:1998wg,GarciaBellido:1998gm,Easther:1999ws,Liddle:1999hq,Finelli:2001db,Bassett:2005xm,Podolsky:2005bw}, were the first signals to be numerically simulated on a 3-dimensional lattice (grid) \cite{Khlebnikov:1997di,Easther:2006vd,Easther:2007vj,GarciaBellido:2007dg,GarciaBellido:2007af,Dufaux:2007pt,Easther:2006gt,Dufaux:2008dn,Price:2008hq}.  Later, processes with multiple scales pushed the limits of numerical simulations to higher resolution, and have employed increasingly sophisticated techniques to maximize the utility of computational resources.

The Laser Interferometer Gravitational Wave Observatory (LIGO) and VIRGO observatories have a peak sensitivities around 100${\rm Hz}$ \cite{Aasi:2013wya}, making them ideal for detecting processes from $\rho_*^{1/4} \sim 10^8\,{\rm GeV}$ for $\zeta=100$ and $\rho_*^{1/4} \sim 10^{10}\,{\rm GeV}$ for $\zeta=0$, which could be associated with models of particle physics beyond the standard model.  On the other hand, the hope for a space-based observatory, the Laser Interferometer Space Antenna (LISA) or its exciting alternative, eLISA \cite{Vitale:2014sla}, has a peak frequency of $10^{-3}-10^{-2}\,{\rm Hz}$ which would correspond to a range from $\rho^{1/4}\sim 10^{3}\,{\rm GeV}$ (for $f_*=10^{-3}$ and $\zeta=100$) to $\rho^{1/4}\sim 10^{6}\,{\rm GeV}$ (for $f_*=10^{-2}$ and $\zeta=1$).  These frequencies could probe signals produced during an electroweak phase transition.  The standard model electroweak phase transition is second-order \cite{Kajantie:1996mn,Laine:1998vn}, however there is some hope that extensions to the standard model would make the electroweak phase transitions first-order  \cite{Carena:1996wj,Delepine:1996vn,Laine:1998qk,Grojean:2004xa,Huber:2000mg,Huber:2006wf,Laine:2012jy,Carena:2008vj,Carena:2012np,Caprini:2006jb,Gogoberidze:2007an,Caprini:2007xq,Kahniashvili:2008pf,Caprini:2009fx}--although the possibility seems less likely as we learn more about the parameters of the electroweak sector \cite{Aad:2012tfa,Chatrchyan:2012ufa}.

Here we work in a more general framework.  The existence of phase transitions in standard model particle physics, as well as the knowledge that descriptions of particle physics beyond the standard model will likely include phase transitions, gives rise to the possibility of new phase transitions with descriptions that are first-order.  Since these transitions are most likely at scales beyond which accelerator experiments can probe, observing the ambient gravitational radiation emitted from these phase transitions is a viable candidate for direct detection of this physics.  Our focus will be independent of an energy-scale, and relevant to a cosmological first-order phase transition in which an order parameter is coupled to an ambient, radiation-dominated Universe.

We work with a model very similar to the the cosmic-fluid-order-parameter-field model \cite{KurkiSuonio:1995vy,Ignatius:1993qn,Hindmarsh:2013xza}, in which a relativistic (perfect) fluid with velocity $U^\mu$ and density $\rho$ is coupled to a scalar field $\phi$.  The field is subject to a canonical kinetic term and a fourth order polynomial potential
\begin{equation}
V(\phi) = \frac{m^2}{2} \phi^2 + \eta \phi^3 + \frac{\lambda}{8} \phi^4
\label{unscaledpot}
\end{equation}
with two minima.  To eliminate the need to set the overall scale of the phase transition, we work in a dimensionless set of variables, \cite{Dunne:2005rt,Giblin:2013kea}
\begin{equation}
\alpha  =  \frac{\lambda m^{2}}{4\eta^{2}},\,\,
\psi  =  \frac{m^{2}}{2\eta}\phi,\,\,
\bar{x}  =  mx,
\end{equation}
along with a dimensionless derivative, $\bar{\partial} = \partial/m$.
Using these, the potential can be parameterized as
\begin{equation}
\label{parampot}
\bar{V}(\psi)=\frac{1}{2}\psi^{2}+\frac{1}{2}\psi^{3}+\frac{\alpha}{8}\psi^{4},
\end{equation}
in which $\alpha$ is the only remaining parameter.  There are two realizations of this potential that are of particular interest.  The $\alpha \rightarrow 1$ limit is the ``thin-wall" limit, where the difference in energy between the true and false vacuum is small, and nucleated bubbles have a radius much larger that the thickness of the bubble walls \cite{Coleman:1977py,Coleman:1980aw}.  This limit is analytically tractable, and so is commonly explored in literature related to false vacuum decay.  Away from this limit, $0 < \alpha < 1$, the difference in energy between the two minima can be large, and the bubble wall thickness is no longer small compared to the bubble radius.

Inspired by \cite{KurkiSuonio:1995vy,Ignatius:1993qn,Hindmarsh:2013xza}, we write the action for a relativistic fluid and field,
\begin{equation}
S = \left(\frac{m^2}{4\eta^2}\right)\int\mathrm{d^{4}}\bar{x}\sqrt{-\bar{g}}\left(\frac{\bar{R}}{2\bar{\kappa}}+\mathcal{\bar{L}}_{\rm fluid} +\mathcal{\bar{L}}_{\phi}\right) ,
\end{equation}
with the fluid Lagrangian
\begin{equation}
\mathcal{\bar{L}}_{\rm fluid} = \bar{p} = \frac{4\eta^2}{m^6}p 
\end{equation}
and the field Lagrangian
\begin{equation}
\mathcal{\bar{L}}_{\phi} = -\frac{1}{2}\bar{\partial}^\mu\psi\bar{\partial}_\mu\psi - \bar{V}(\psi).
\end{equation}
The gravitational piece of the action can also be written in a dimensionless way,
\begin{equation}
\bar{R} = \frac{R}{m^2},
\end{equation}
where
\begin{equation}
\bar{\kappa} = \frac{m^4}{4\eta^2}\kappa.
\end{equation}
As in \cite{Schutz:1970my}, we can start to construct equations of motion for the fluid using the stress-energy tensor for the fluid,
\begin{eqnarray}
\label{fluidset}
\bar{T}^{\mu\nu}_{\rm fluid} & = & (\bar{\epsilon}+\bar{p})U^{\mu}U^{\nu}+\bar{p}\bar{g}^{\mu\nu}.
\end{eqnarray}
In the absence of a coupling to other constituents, $\bar{T}^{\mu\nu}_{\rm fluid}$ is conserved,
\begin{equation}
\label{fluidcset}
\bar{D}_{\mu}\bar{T}^{\mu\nu}_{\rm fluid} = 0,
\end{equation}
a statement from which one can extract equations of motion for the fluid.  The field has an analogous set of equations, with
\begin{equation}
\label{phiset}
\bar{T}^{\mu\nu}_{\psi} = \bar{\partial}^\mu \psi \bar{\partial}^\nu \psi - \bar{g}^{\mu\nu}\left( \frac{1}{2}\bar{\partial}_\alpha\psi\bar{\partial}^\alpha\psi + \bar{V}(\psi)  \right) ,
\end{equation}
so that
\begin{equation}
\label{phicset}
\bar{D}_{\mu}\bar{T}^{\mu\nu}_{ \phi} = 0.
\end{equation}
In this model, the two are coupled using a phenomenological diffusive term,
\begin{equation}
j^\nu = \xi U^\mu \bar{\partial}_\mu\psi \bar{\partial}^\nu \psi,
\label{source}
\end{equation}
which we add to Eq.~\ref{phicset} and subtract from Eq.~\ref{fluidcset} so that total stress-energy remains conserved \cite{KurkiSuonio:1995vy,Ignatius:1993qn}.

Given an equation of state for the fluid, $\bar{p} = w \bar{\epsilon}$, we can write the equations of motion for the field,
\begin{equation}
\label{fieldEOM}
\bar{\Box} \psi = \frac{\partial \bar{V}}{\partial \psi} + \xi U^\mu \bar{\partial}_\mu \psi,
\end{equation}
and for the fluid,
\begin{equation}
\label{spaceeom}
\begin{split}
\bar{\partial}_{t}U^{i} =& \frac{wU^{t}U^{i}}{\left(1+(1-w)U^{k}U_{k}\right)} \\
& \times \left( \bar{\partial}_{k}U^{k}-\frac{U_{j}}{\left(U^{t}\right)^{2}}\left(U^{k}\bar{\partial}_{k}U^{j}+\frac{w}{1+w}\bar{\partial}^{j}\ln\left(\bar{\epsilon}\right)\right)\right) \\
& - \frac{1}{U^{t}}\left(U^{k}\bar{\partial}_{k}U^{i}+\frac{w}{1+w}\bar{\partial}^{i}\ln\left(\bar{\epsilon}\right)\right)+J^{i}
\end{split}
\end{equation}
and \footnote{We correct a sign error that appears in Eq.~24 of \cite{Giblin:2013kea}.}
\begin{equation}
\label{energyeom}
\begin{split}
\bar{\partial}_{t}\ln\left(\bar{\epsilon}\right) =& -\frac{(1+w)U^{t}}{\left(1+(1-w)U^{k}U_{k}\right)} \\
& \times \left(\frac{1-w}{1+w}U^{i}\bar{\partial}_{i}\ln\left(\bar{\epsilon}\right)+\bar{\partial}_{i}U^{i}-\frac{U_{j}U^{k}}{\left(U^{t}\right)^{2}}\bar{\partial}_{k}U^{j}\right) \\
& -\left(1+w\right)\frac{U^{i}J_{i}}{\left(U^{t}\right)^{2}} - \frac{1}{U^{t}}U^{\mu}\frac{j_{\mu}}{e^{\ln\left(\bar{\epsilon}\right)}}
\end{split}
\end{equation}
where
\begin{equation}
\begin{split}
J^{i} =& \frac{1}{e^{\ln\left(\bar{\epsilon}\right)} U^{t}}\left(\delta_{j}^{i}+\frac{w}{\left(1+(1-w)U^{k}U_{k}\right)}U^{i}U_{j}\right) \\
& \times \left(\frac{j^{j}}{1+w}+U^{j}U^{\mu}j_{\mu}\right).
\end{split}
\end{equation}
The phenomenological coupling $j^\mu$ between the fluid and field allows the latent heat energy of the field to be converted into bulk motions of the fluid.  Such a term term is diffusive in nature, so a stronger coupling constant $\xi$ deposits more energy (and entropy) into the fluid and, at the same time, slows down the bubble walls \cite{KurkiSuonio:1995vy,KurkiSuonio:1996rk,Giblin:2013kea}  (See \cite{Giblin:2013kea} for more details of this computational model.)  We take $w=1/3$ for the entirety of this work.

We do want to point out that there is an instability in this model; since the phenomenological $j^\mu$ enters into the equations of motion as a derivative coupling, it is possible for a phase transition with large field gradients and fluid velocities to enter a regime in which modes can grow exponentially.  The time component of the coupling is always a ``normal looking" friction term, $\xi U^0 \partial_0 \psi $, but the spatial term, $\xi U^i \partial_i \bar{\psi}$, can  dominate the dynamics, and can do so with the ``wrong" sign.  In practice we see this occur for large values of $\xi$ and large field gradients, but the simulations break down when they enter these regimes.  In this work we will focus only on those regions of parameter space in which our model is stable.

In recent years there has been great progress toward fully understanding first-order phase transitions in the presence of a fluid.  Analytic work sets up the expectation that bubble wall velocities are affected by fluids \cite{KurkiSuonio:1995vy,Ignatius:1993qn,Hindmarsh:2013xza}, and analytic and numerical work sets some expectations for the wall velocity \cite{Steinhardt:1981ct,Kamionkowski:1993fg,Huber:2011aa,Huber:2013kj}.

This has culminated recently in a high-resolution numerical study \cite{Hindmarsh:2013xza}, in which the authors calculate the gravitational radiation for a specific first-order phase transition and carefully study the relative contributions of the fluid and field source.  In this work we extend this analysis to a wider range of scenarios; looking at transitions over a wide range of parameters.  We also set expectations for the amplitude of the gravitational wave signal that can be used by the observational community.

In Section \ref{bubnuc}, we first review and apply a particular model of bubble nucleation in a first-order phase transition.  We then review the formalism behind gravitational wave generation in Section \ref{gwgen}.  Finally, in Section \ref{compute}, we describe the numerical procedure we use for calculating the gravitational wave spectrum and present our results in Section \ref{results}.

\subsection{Notation}

Since this work represents the convergence of independent lines of investigation, our choice of notation represents a best-attempt to reconcile the choices made by previous authors.  In this work we use $\alpha$ to refer only to the dimensionless parameter that appears in Eq.~\ref{parampot} \cite{Dunne:2005rt,Giblin:2013kea}; a parameter that is absent in most other models.  Due to the parameterization of our potential, we use $\beta$ to represent the ratio of the energy in the scalar field to the energy in the fluid, $\rho_{\phi}/\rho_{\rm fl}$, where we approximate the total energy density, $\rho = (1-\beta) \rho_{\rm fl} \approx \rho_{\rm fl}$.  This is consistent with \cite{Giblin:2013kea} but inconsistent with the definition used in \cite{Kamionkowski:1993fg,Child:2012qg} and others, where this ratio is called $\alpha$.  It is also related to the convention in \cite{KurkiSuonio:1995vy,Ignatius:1993qn,Hindmarsh:2013xza} where this ratio is called $\alpha_T = 4 \beta$, with the factor of 4 translates between the initial energy in the field andthe latent heat of the potential. This should also not be confused with the parameter $\beta^{-1}$ that is commonly used \cite{Kamionkowski:1993fg,Child:2012qg} to denote the timescale of the transition, which we discuss in Sec.~\ref{bubnuc}.

We will use asterisk subscripts, $_*$, to denote physical quantities at the time of the phase transition, especially in cases where there is ambiguity between the values of these quantities at the time of the transition and today.  When using Fourier-transformed quantities, we explicitly include the $\vec{k}$ dependence--and leave off the tilde as a redundant markingOur Fourier convention is
\begin{equation}
A(\vec{k}) =  \int d^3x \,A(\vec{x}) e^{2\pi\,i\vec{k}\cdot \vec{x}}.
\end{equation}
Of course, in practice we use Fast Fourier Transforms, and employ the {\tt FFTW} package for this procedure \cite{FFTW05}.

\section{Initial Conditions}
\label{bubnuc}

Now that the evolution of a single bubble is well understood, we aim to tackle the full problem of colliding bubbles during a phase transition.  In order to do so, we first need to determine initial conditions in our simulation.  Each bubble is initialized with a profile determined by a numerical ``shooting" method, as in \cite{Giblin:2013kea}.  This approximates the instanton solution for a given $\alpha$ where the initial bubble radius is given by \cite{Dunne:2005rt}
\begin{equation}
\bar{R_0} = m R_0 \approx (1-\alpha)^{-1}.
\end{equation}

For phase transitions, we also need to determine where to put the bubbles on our initial slice.  In principle there are several ways that can be used to determine the false vacuum decay rate rate.  One method is to directly calculate the quantum effective action as in \cite{Dunne:2005rt} or \cite{Enqvist:1991xw}, while another approximate method is to assume a slowly varying potential and an exponential nucleation rate as in \cite{Turner:1992tz}.

The latter of these methods appears more often in the literature.  In the absence of a true temperature-dependent potential, we use this and simply leave the nucleation rate as a free parameter, similar to \cite{Child:2012qg}.  We note that, as in that work, the inter-bubble spacing $l_*$ will be related to the nucleation rate, so that given a number of bubbles per hubble volume $n$, we have
\begin{equation}
l_* \sim 0.55 n^{-1/3} H_*^{-1} .
\end{equation}
Typically $l_*$ is taken to be small enough so that there will not be large inhomogeneities in the present Universe, $l_* \ll H_*^{-1}$.  The expected value of $l_*$ is generically of order $10^{-2}$ \cite{Kamionkowski:1993fg}, although studied values have ranged up to $l_* \sim 10^{-1} H_*^{-1}$ \cite{Child:2012qg}.  In the thin-wall, $\alpha \rightarrow 1$, vacuum bubble, $\xi=0$, limit, the bubble walls accelerate to a speed of light very quickly, and so the inter bubble spacing is also a time scale that determines how long the phase transition lasts.  When the coupling is non-negligible, $\xi > 0$, we have to be more careful to define a time-scale.  Since we understand the terminal wall velocity, it's attractive to take this time-scale to be $l_*/v_f$, where $v_f$ is the terminal wall velocity given a set of $\alpha$, $\beta$ and $\xi$.

\begin{figure}[hbt]
  \centering
    \includegraphics[width=0.45\textwidth]{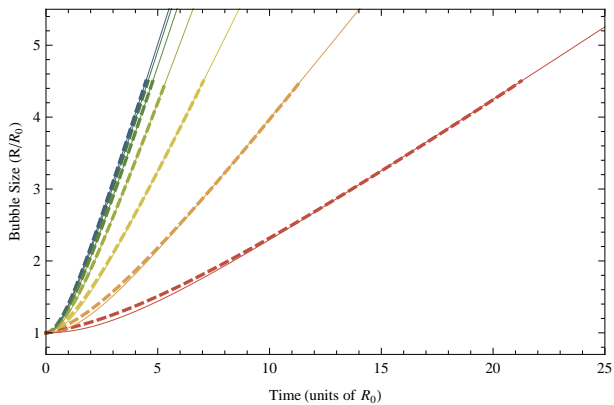}
  \caption{\label{hyperbolicapproximation}
Here we plot the size of bubbles in the fully non-linear simulations (dashed line) as in \cite{Giblin:2013kea} compared to the hyperbolic approximation (thin solid lines).  The consistency of the hyperbolic approximation justifies the use of Eq.~\ref{rad_vs_time}.  This is a sample for $\alpha = 0.45$, $\beta=0.028$ and varying couplings $\xi$.  From top to bottom the couplings are $0.1$, $0.2$, $0.4$, $0.8$, $1.6$, $3.2$, $6.4$.}
\end{figure}
The actual story is more complicated, in the case of small $l_*$ or large $\xi$, it is not always true that the bubbles reach their terminal velocities before colliding.  Empirically we see that the bubble growth is approximately hyperbolic, described by the terminal wall velocity and an additional scale $r_s$ that determines how quickly the wall reaches this velocity,
\begin{equation}
\label{rad_vs_time}
R(t)/R_0 \approx \sqrt{r_s^2+(v_f t)^2} + 1 - r_s .
\end{equation}
Eq.~\ref{rad_vs_time} is an exact relationship in the case of vacuum bubbles where $r_s=1$ and $v_f=1$.  Fig.~\ref{hyperbolicapproximation} shows that for nonzero couplings, this is a good approximation for much of the life of these bubbles, especially for smaller couplings, or at later times once the bubbles have reached their final velocity.  For each value of $\xi$, we find a best fit for $r_s$ and $v_f$ using the wall position after it has grown past $2 R_0$, but before it collides with itself at the simulation boundary.

\section{Gravitational Wave Generation}

\label{gwgen}

Gravitational radiation is energy stored in propagating metric perturbations.  To define these degrees of freedom we employ synchronous gauge, and express perturbations, $h_{ij}^{\rm TT}$ as
\begin{equation}
ds^2 = dt^2 - \left[\delta_{ij} + h_{ij}^{\rm TT}\right]dx^idx^j,
\end{equation}
where the superscript $^{\rm TT}$ implies that we have imposed the transverse-traceless conditions,
\begin{equation}
{h^{\rm TT}}^{i}_{i} = 0,\,\,\partial^i h^{\rm TT}_{ij} = 0.
\end{equation}
Theoretical work generally expresses the power in stochastic gravitational radiation using the quantity 
\begin{equation}
\Omega_{GW,0}h^{2} = \frac{1}{\rho_c}\frac{d \rho_{GW}}{d\ln k}
\label{gwexp}
\end{equation}
where $\rho_{GW}$ is the energy in gravitational waves. The energy density of gravitational waves stored in the metric perturbations is
\begin{equation}
\frac{d\rho_{GW,*}}{d\ln k}=\frac{m_{pl}^{2}k^{3}}{32\pi}\frac{1}{V}\underset{i,j}{\sum}\int\mathrm{d}\Omega\left|\dot{h}_{ij}^{TT}(\vec{k},t)\right|^{2}\,.
\end{equation}

The process of calculating Eq.~\ref{gwexp} is straightforward; the stress-energy tensor $T_{\mu\nu}$ for a process that might source gravitational waves is projected onto the transverse-traceless anisotropic stress tensor,
\begin{equation}
S_{ij}^{\rm TT} = P_{ik}\left(T_{kl} - \frac{\delta_{kl}}{3}T_i^i\right)P_{kj}
\end{equation}
where the projection operator is $P_{ij} = \delta_{ij} - k_ik_j/k^2$.
There are two contributions to the stress-energy:  (1) from the field, Eq.~\ref{fluidset}, and (2) from the fluid, Eq.~\ref{phiset}.
As the simulation progresses we can calculate these two sources, and use them to evolve the metric perturbations, which obey the sourced Klein-Gordon equation,
\begin{equation}
\label{evolutioneq}
\left(\frac{d^{2}}{d\bar{t}^{2}}+\bar{k}^{2}\right)\bar{h}^{\rm TT}_{ij}(\bar{k},\bar{t})=2\bar{\kappa}\bar{S}^{\rm TT}_{ij}(\vec{k},t).
\end{equation}
In practice, we perform a more (numerically) efficient calculation. We first absorb the overall constant into  $\bar{b} \equiv \bar{h}/2\bar{\kappa}$.  Also, rather than projecting the source at every step, we only project the metric perturbations into transverse-traceless space, $\bar{b}_{ij}^{TT}$, when we calculate observable quantities.  We lastly do not calculate pure-trace terms in the source, using the expression
\begin{equation}
\label{pseudosrc}
T_{ij}^{\rm T} = \bar{\partial}_{i}\psi\bar{\partial}_{j}\psi + (\bar{\epsilon}+\bar{p})U_{i}U_{j}.
\end{equation}
as a source instead.  This is an equivalent process algebraically, but we also justify its numerical use in Sec.~\ref{compute}.

This approach allows us to track all of the (possible) modes on our finite lattice so that at any time we can calculate the power spectrum of gravitational radiation.   At that point we calculate
\begin{equation}
\Omega_{\rm GW,*}(\vec{k}) = 3n^{2}\left(\frac{\beta}{\Delta\bar{V}}\right)^{2}\frac{\bar{k}^{3}}{\bar{L}^{5}}\int\mathrm{d}\Omega\left|\dot{\bar{b}}_{ij}^{TT}(\vec{k},\bar{t})\right|^{2}.
\end{equation}
We can then use the known transfer functions to calculate the gravitational wave power today \cite{Easther:2006gt,Easther:2007vj}
\begin{equation}
\Omega_{\rm GW,0}h^{2}=\Omega_{r,0}h^{2}\left(\frac{g_{0}}{g_{e}}\right)^{1/3}\Omega_{\rm GW,*}(\vec{k})\,.
\label{heighttransfer}
\end{equation}
Here we have $\Omega_{r,0} h^2 = 4\times 10^{-5}$ \cite{Dodelson:2003ft} (assuming massless neutrinos) and the ratio of relativistic degrees of freedom today, $g_0$, compared to relativistic degrees of freedom at matter-radiation equality, $g_e$, is $g_{0}/g_{e} = 1/100$--although our predictions rely only very weakly on this factor.
Our simulation volume has size $L^{3}=(H_*^{-1}/2)^{3}$, where $H_*^{2}=\kappa\rho_{fl}/3$ for a flat Universe. Lastly, the discrete quantities $\bar{k}$ are given in integer multiples of $d\bar{k}=4\pi H/m $, where $m$ is the potential parameter that will set the physical scale.  Although we will not transfer the frequencies, it is easy to do using Eq.~\ref{freqestimator}.
We thus have that the overall amplitude of the spectrum is independent of the scale at which the transition occurred, while the frequency will depend on the scale.

\section{Numerical Method}
\label{compute}

\begin{figure*}[htb]
\includegraphics[width=0.95\textwidth]{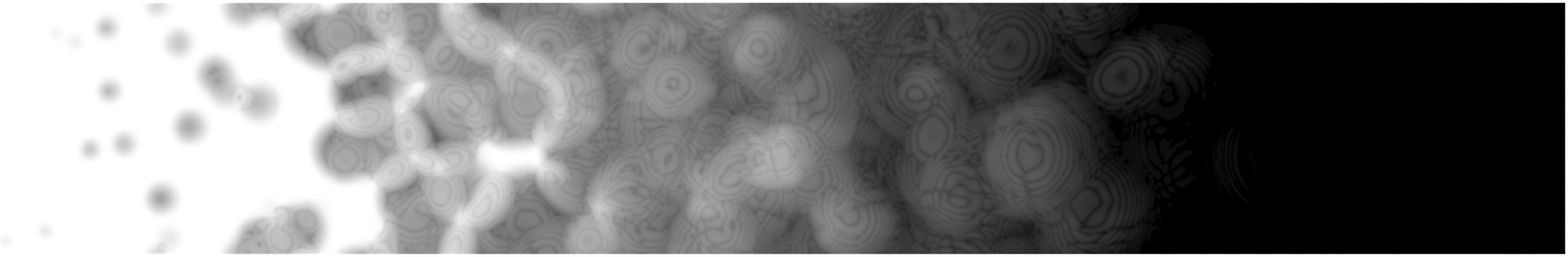}
\includegraphics[width=0.95\textwidth]{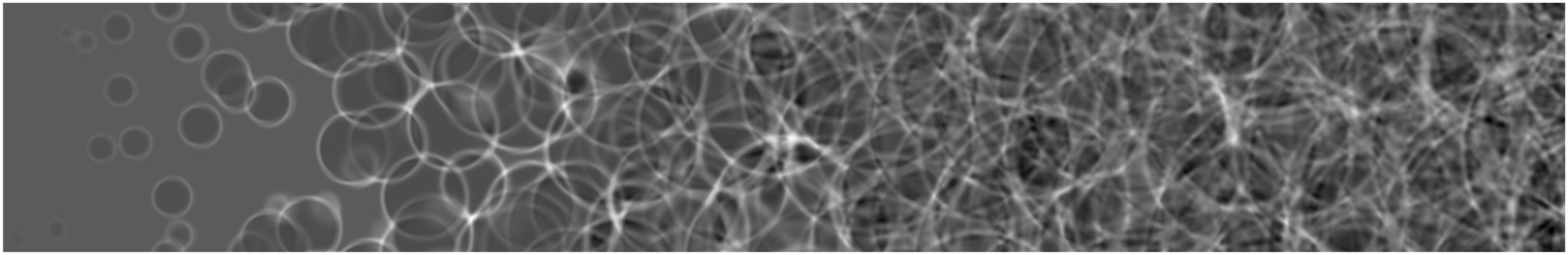}
\caption{\label{prettyevolution} A time-depedent representation of our simulations.  The vertical axis, $y$, is a one-dimensional slice through the box, and the horizontal axis is a space-like hyper surface, $x+ c_f t$, for some $c_f<1$ that spans the course of a simulation.  The upper panel shows a scaled version of the field value, roughly $ln|\psi-\psi_{-}|$, with $\psi_{-}$ the field value in the true minimum.  The lower panel shows the logarithm of the energy density of the fluid $\ln (\epsilon)$.  Lighter regions are at higher field values/energy density.  The scale is exaggerated to show definition.  The reader should notice that the fluid is being pushed outward by the field after the initial nucleation of bubbles and continues to ``slosh" around after the field has homogenized about the true minimum.}
\end{figure*}

We use the same method of integration as we did in \cite{Giblin:2013kea} in order to evolve the field, Eq.~\ref{fieldEOM}, and fluid, Eq.~\ref{spaceeom}.  We then use the algorithm from \cite{Child:2012qg} to evolve the metric perturbations.  Since the equations of motion for the metric perturbations are linear,  we evolve them in Fourier space using a 4th-order Runge-Kutta integrator.    For all simulations, we use a timestep $dt = dx/10$.  However, we only evolve the metric perturbations at every other step, effectively evolving the metric with a timestep $dt = dx/5$.  We find the fluid and metric evolution to take roughly 20\% of the computational time.  Calculating the Fourier transforms for every component of the stress-energy tensor takes the remaining (80\%) of the computational time.  Fig.~\ref{prettyevolution} shows a time-like hyper-surface that cuts through our four-dimensional simulation.

\begin{figure}[tb]
  \centering
    \includegraphics[width=0.45\textwidth]{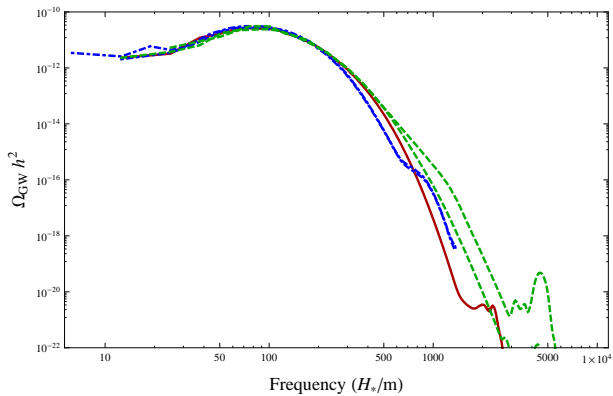}
  \caption{\label{a.45res_comp}Agreement of spectrum position and amplitude for $\alpha = 0.45$ for various resolutions and box sizes.  All spectra have different initial conditions with $l_* = 5.5$, $\beta = 0.1$, and $\xi=3.2$.  The solid red line is $N^3 = 256^3$ and $L=36 R_0$. The dot-dashed blue lines are $N^3 = 256^3$, $L=72 R_0$; and $N^3 = 128^3$, $L=32 R_0$.  The dashed green lines are $N^3 = 512^3$, $L=36 R_0$; and $N^3 = 256^3$, $L=18 R_0$.  Spectra are plotted during the coalescence phase when $R(t) = l_*$, and the spectra are scaled so that the solid red line has $L = 2 H_*^-1$.  Disagreement begins around $500 H_*/m$.  The bump at very high frequencies is a numerical artifact. }
\end{figure}

When performing the integration, we found our simulations to be more stable for smaller values of $\beta$, especially in the thin-wall, $\alpha=0.96$, case.  Due to this, we opt to use $\beta = 10^{-2}$, $\beta = 0.028$ (used in \cite{Hindmarsh:2013xza}, and $\beta = 0.1$ (for illustrative purposes) for the $\alpha = 0.45$ case and $\beta=10^{-3}$ for the $\alpha=0.96$ case.  The relatively thick walls in the $\alpha=0.45$ case allow us to use a lower resolution ($256^3$) without loss of numerical accuracy.  To test this, we ran a fiducial model for a variety of resolutions and initial conditions, as shown in Fig.~\ref{a.45res_comp} and saw that the spectra all peak at the same amplitude and frequency, and the numerical parameters only had an effect on numerical artifacts (mostly at high frequency).
We also compared the differences in our spectra at $\alpha = 0.96$ for different resolutions ($N=512^3$ and $N = 678^3$) and found similar agreement between spectra near the peak, even at lower resolutions where the wall is not resolved as well.  

\begin{figure}[tb]
  \centering
    \includegraphics[width=0.45\textwidth]{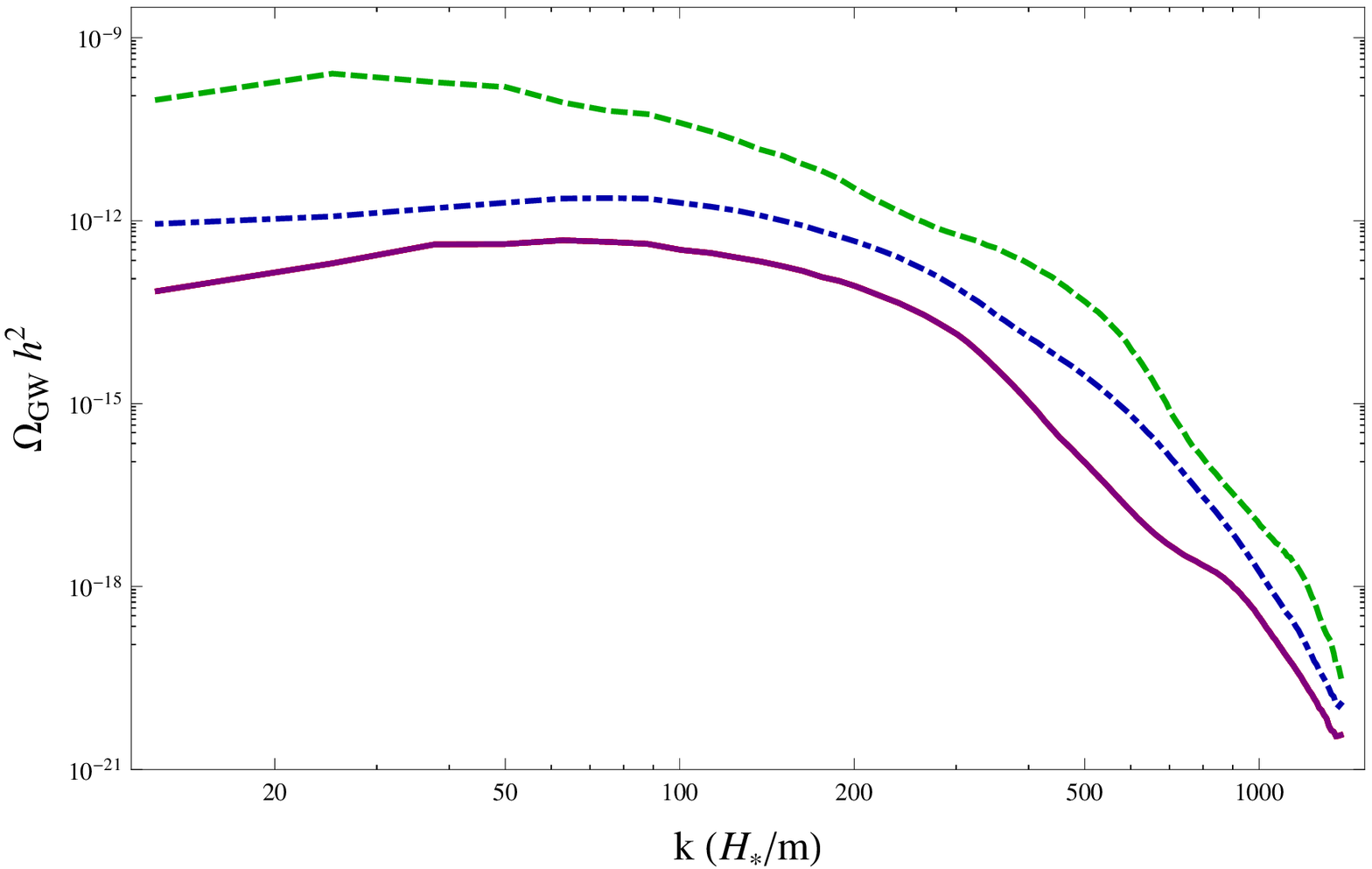}
  \caption{\label{projections}Equivalence of gravitational wave spectra with the transverse-traceless projection taken at different points in the code.  The dashed line (green) is a result of making no projections at all, and using the full $T_{ij}$ as the source in Eq.~\ref{evolutioneq}.  The dot-dashed (blue) line is a result of sourcing using $T_{ij}^{\rm T}$ (Eq.~\ref{pseudosrc}) and performing no further projections.  There are three solid lines, indicating different projected versions of the spectra: using the projected anisotropic stress tensor $S^{\rm TT}_{ij}$ as a source and not projecting the metric perturbations (dark red), using the full $T_{ij}$ as the source and projecting the metric perturbations $\bar{b}_{ij}$, (light red), and finally using $T_{ij}^{\rm T}$ and projecting the metric perturbations $\bar{b}_{ij}$ (light purple).  The projected lines are virtually indistinguishable.  This was for a simulation in which $\alpha = 0.45$, $\beta = 0.028$, $L = 36$, $N^3 = 128^3$, $\xi = 3.2$, and $l_* = 6.3$, where the spectra are evaluated at a Hubble time, $t = H_*^{-1}$. }
\end{figure}

We also briefly examine the effect of projecting the metric perturbations using different schemes.  Because only the transverse-traceless
part of the metric perturbations will contribute to the gravitational wave spectrum, it is always necessary to project onto this space. In Fig.~\ref{projections} we show that the algebraic equivalent choices of projection are robustly equivalent in the numerical method.  For this reason we do the most computationally efficient process and source Eq.~\ref{evolutioneq} with $T_{ij}^{\rm T}$ and project only when calculating the gravitational wave power spectrum -- note that it is easier to calcite $T_{ij}^{\rm T}$ than it is to calculate $T_{ij}$.

\section{Results}
\label{results}
 
We will examine the two $\alpha$ cases in turn, beginning with the $\alpha = 0.45$ case in Sec.~\ref{45case} and the $\alpha=0.96$ case in Sec.~\ref{96case}.  We give our simulations a volume $L^{3} = (H_*^{-1}/2)^{-3}$, where $H_*$ is the Hubble rate during the process.  We chose this box size to optimally balance the desire for resolution on scales smaller than $H_*^{-1}$ while also minimizing any effects from periodic boundary conditions (which set in when the simulation runs for much longer than the light-crossing time of the box $t> L$).  We also acknowledge the fact that neglecting Hubble friction limits our final time to less than the Hubble time $t < H_*^{-1}$, at which point expansion effects become extremely important.

We express the gravitational wave spectrum frequency in terms of the physical Hubble scale at the time of the transition, $H_*$, and the mass scale $m$ in the potential of the problem.  Given that $\bar{H}_* = H_* / m$, and our Fourier transform returns units of angular frequency, $dk = 4\pi H$, we plot the frequency in units of $\left[k\right]= H_*/m$.  We do not scale the amplitudes of the gravitational wave spectra, as the energy scale $\bar{H}_*$ does not appear in our final expression for the amplitude.

In order to compare spectra at equal points in their evolution, we plot the spectra as a function of the bubble radius, $R(t)$, given by Eq.~\ref{rad_vs_time} rather than as a function of time.  This helps to preserve the intuition that the time-scale of the phase transition is what helps determine the observational consequence.

\subsection{The thick-wall, $\alpha=0.45$ case}
\label{45case}

Unless otherwise stated, all simulations for the $\alpha=0.45$ case are in a volume $L = 36 R_0$ with $N_b = 32$ bubbles.  Therefore the mean bubble spacing is 
\begin{equation}
l_* = 0.55 (N_b)^{-1/3} L = 6.28 R_0.
\end{equation}
Once the bubbles reach half this size (although really somewhat less, since some bubble centers are closer than $l_*/2$ and the walls are thick for this choice of $\alpha$), they begin to collide.

\begin{figure}[htb]
  \centering
    \includegraphics[width=0.45\textwidth]{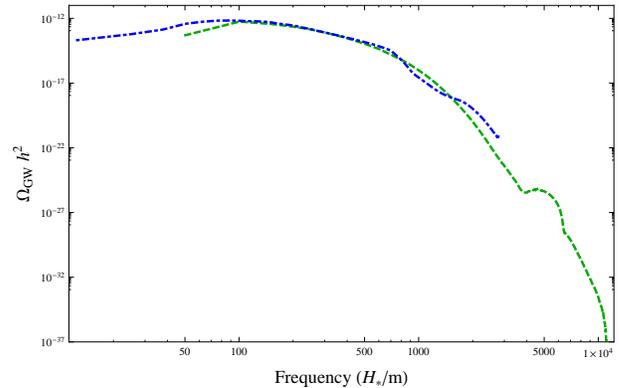}
  \caption{\label{hh_comp} A spectrum run with parameters similar to that of \cite{Hindmarsh:2013xza}, for a bubble spacing $l_* \sim 7 R_0$, coupling $\xi = \eta/m = 3.2$, $\beta = \alpha_{T}/4 = 0.028$, and $\alpha = 0.45$.  We run in a box with $N^3 = 256^3$, at two scales, $L = 72 R_0$ (dot-dashed, blue) and $L = 18 R_0$ (dashed, green).  The amplitudes are scaled so that the latter is half of a Hubble volume.  The spectra are plotted at a time equal to the largest amplitude spectrum shown in \cite{Hindmarsh:2013xza}, around $t = 1400 T_c^{-1}$ (in the units of \cite{Hindmarsh:2013xza}), or approximately $18.6 R_0 ~ 2.8 l_*$. }
\end{figure}

It is important for us to compare our numerical method with the results of Hindmarsh et al \cite{Hindmarsh:2013xza}.  To make this comparison, we run our code with approximately equivalent parameter choices, both at a high resolution ($dx = 0.07 R_0$, comparable to \cite{Hindmarsh:2013xza})  although in a smaller volume; and at a lower resolution ($dx = 0.28 R_0$) with a larger volume.  The result of these simulations can be seen in Fig.~\ref{hh_comp}.  We find a similar bubble wall terminal velocity, $v_f = 0.4$, and a similar spectral shape.  There still remain some differences in our models, however.  In \cite{Hindmarsh:2013xza} bubbles are nucleated at different times during the first stages of the simulations, but with approximate numerical profiles.  We also use the transfer function, Eq.~\ref{heighttransfer}, to express the power spectrum, and we find that $\Omega_{GW} h^2 \sim 10^{-13}.$

\begin{figure*}[p]
    \begin{tabular}{rcc}
      & $R(t) = l_*$ & $R(t) = 2.5 l_*$ \\
\\
      \rotatebox{90}{ \hspace{10 mm} $\beta = 0.01$} \hspace{5 mm}
      & \includegraphics[width=0.45\textwidth]{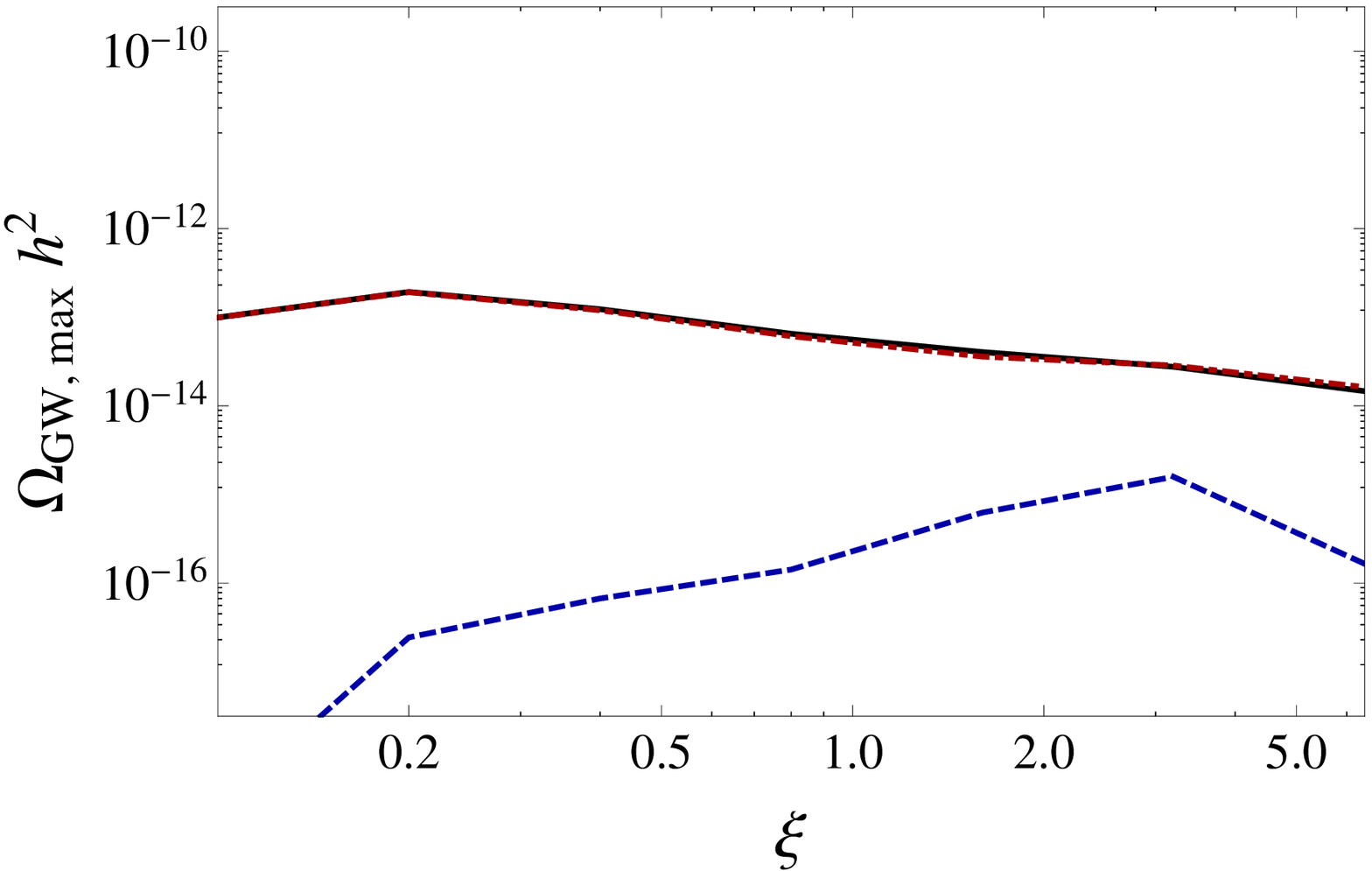}
      & \includegraphics[width=0.45\textwidth]{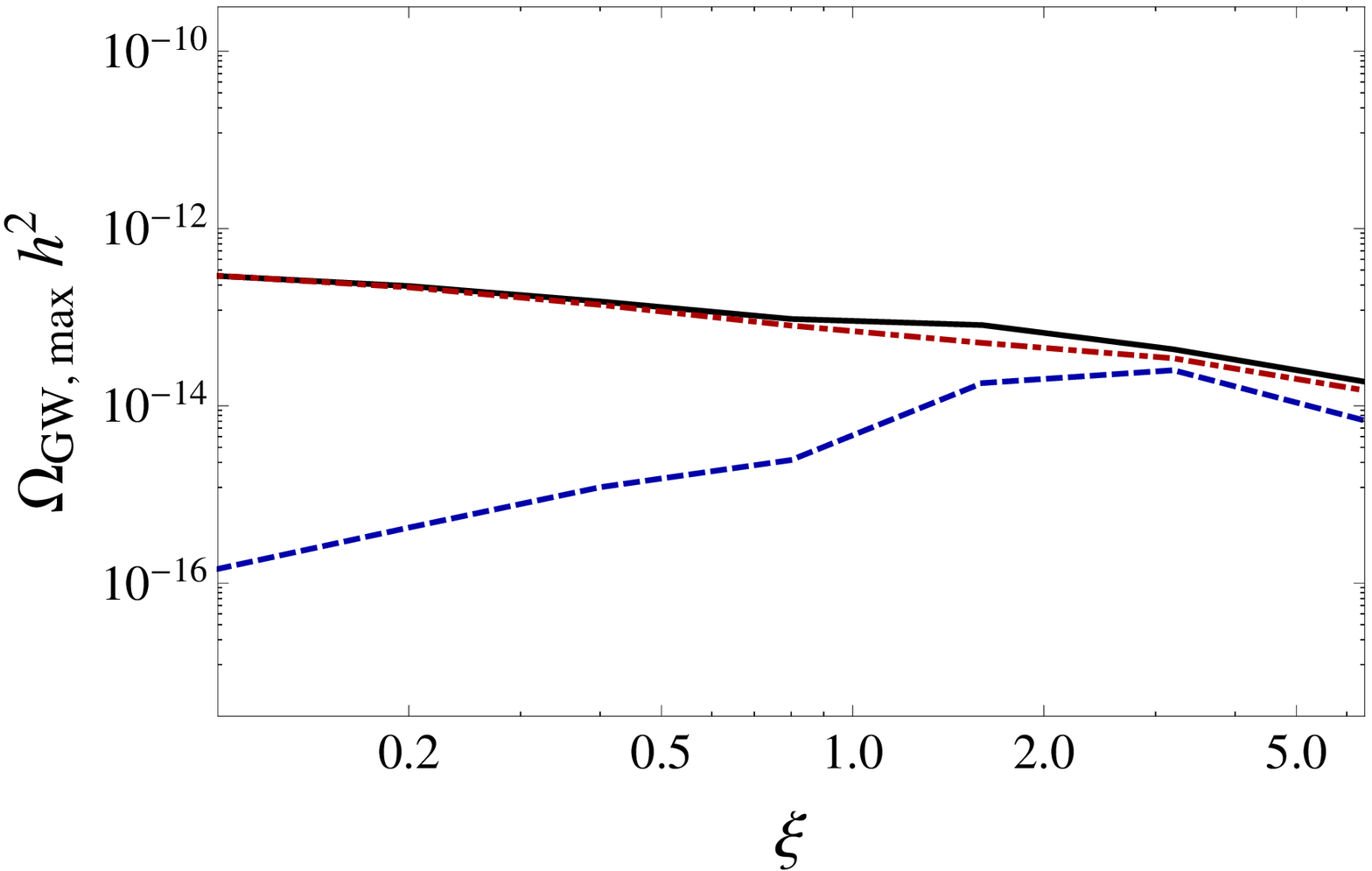} \\
\\ \\ 
      \rotatebox{90}{ \hspace{10 mm} $\beta = 0.028$}  \hspace{5 mm}
      & \includegraphics[width=0.45\textwidth]{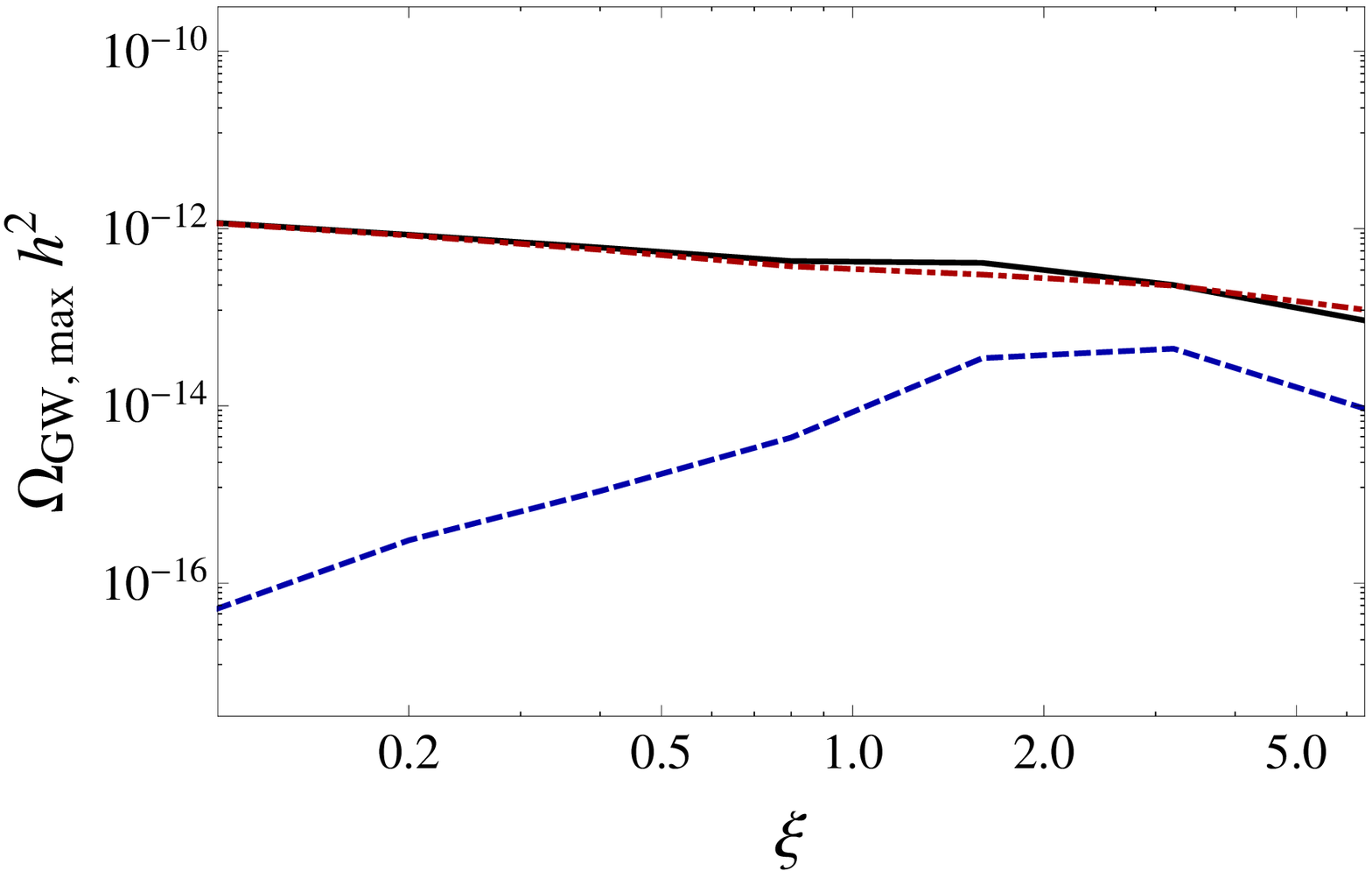}
      & \includegraphics[width=0.45\textwidth]{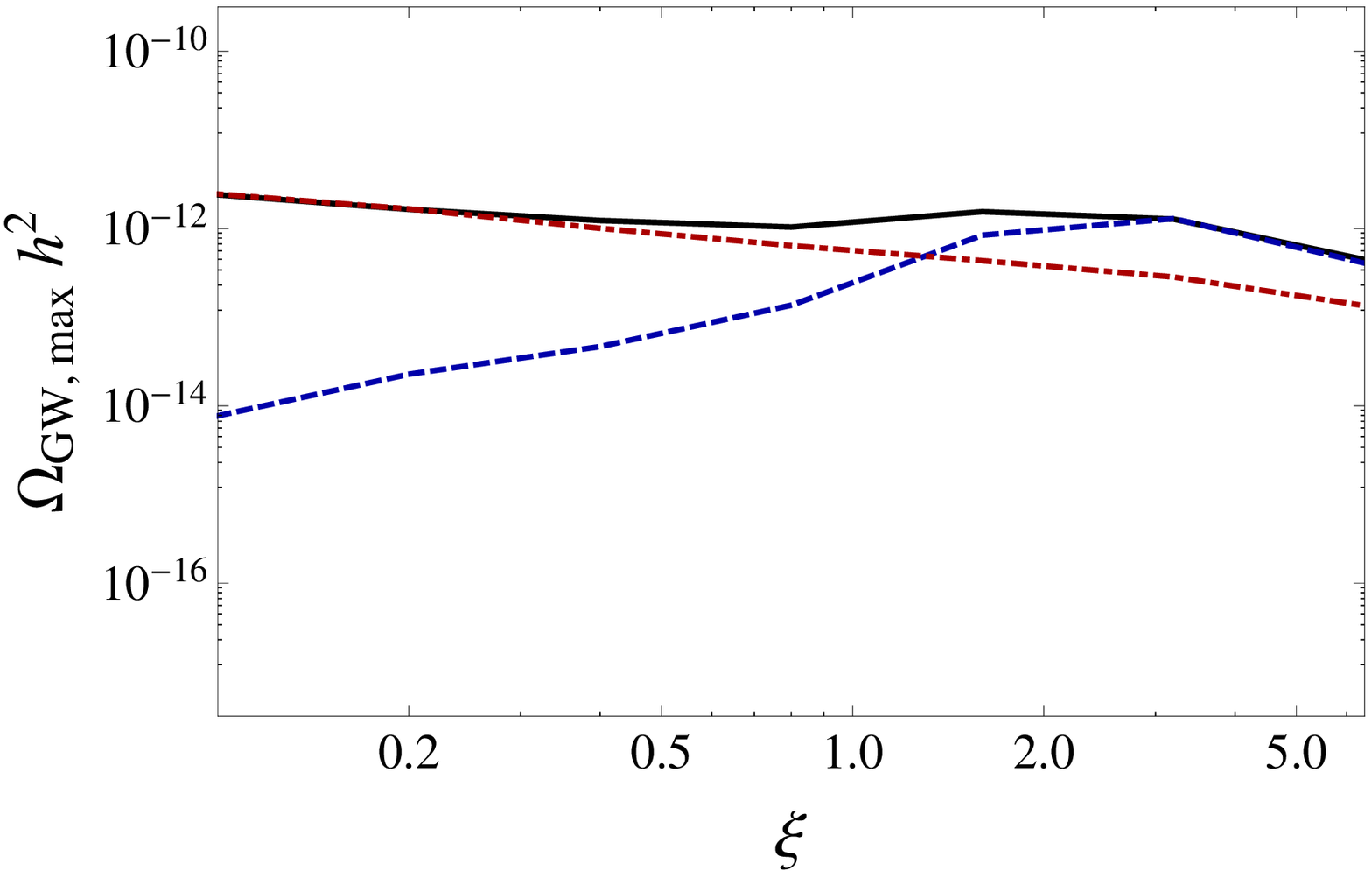} \\
\\ \\
      \rotatebox{90}{ \hspace{10 mm} $\beta = 0.1$}  \hspace{5 mm}
      & \includegraphics[width=0.45\textwidth]{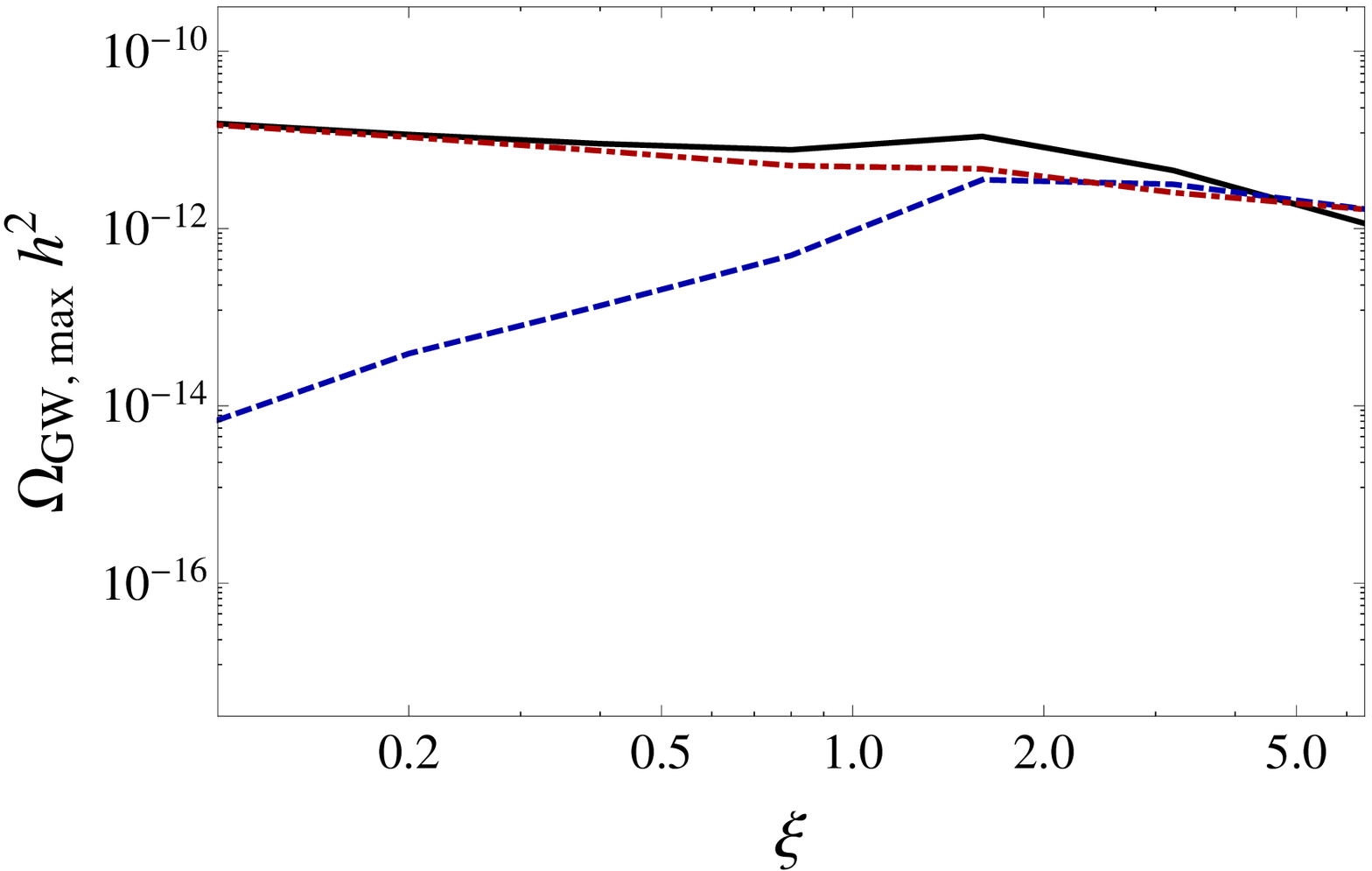}
      & \includegraphics[width=0.45\textwidth]{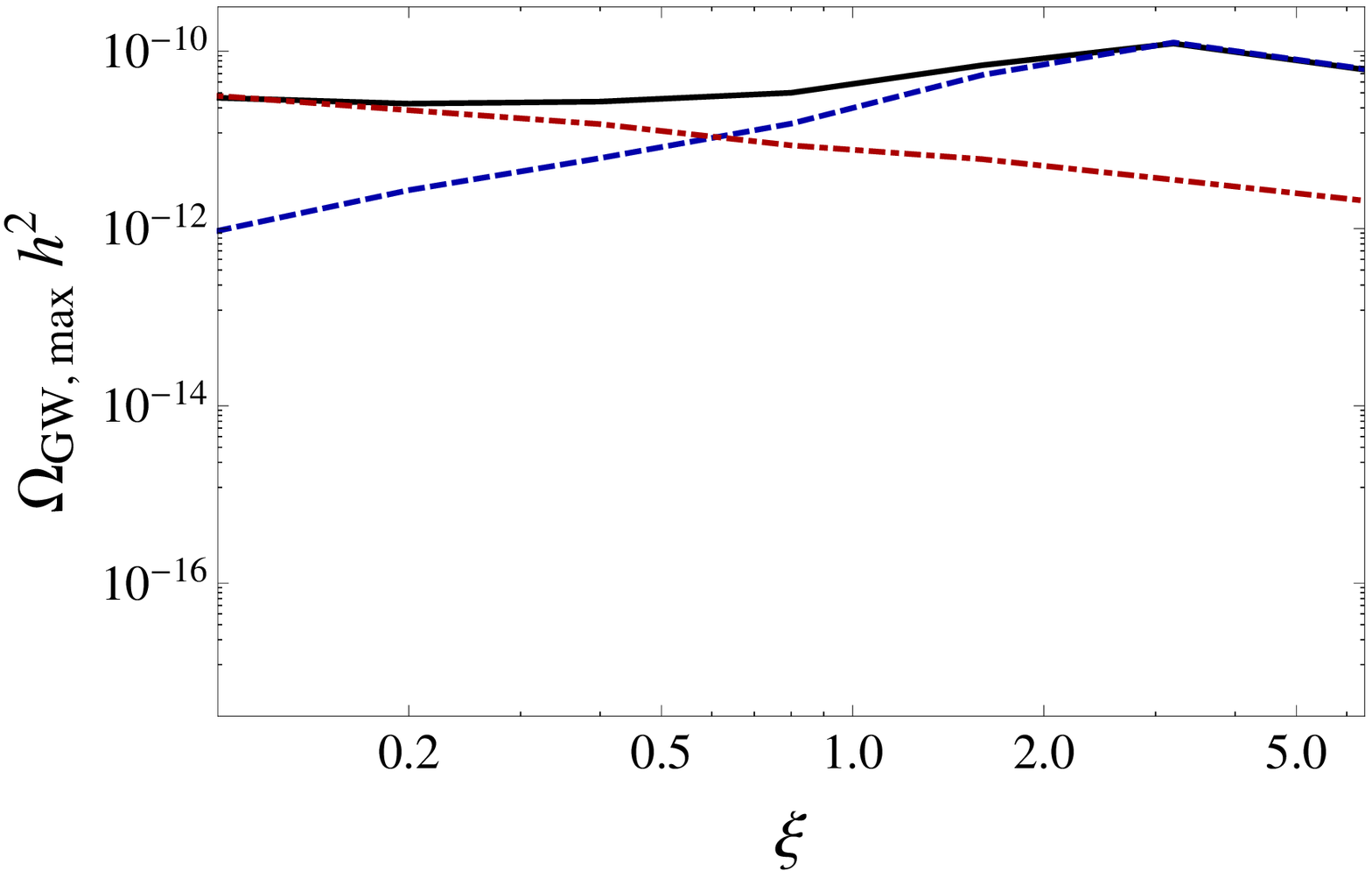} \\
\\
    \end{tabular}

        \caption{Plots of amplitudes vs coupling for varying $\beta$ values and $\alpha = 0.45$.  All couplings plotted are of the form $\xi = 0.1 \times 2^n$.  The solid black line indicates the total power spectrum, red dot-dashed the field contribution, and blue dashed the fluid.  The time required to evolve the simulations to $R = 2.5 l_*$ exceeded a Hubble time for the highest couplings, $1.6$, $3.2$, and $6.4$, likely making the presented amplitudes upper bounds for these couplings.  \label{fig_0.45amplitude_table}}
\end{figure*}

This model is only one example in a complex and diverse parameter space.  To address the relative (and overall) contributions to the gravitational wave spectrum, we want to characterize this parameter space more generally.  For this we will take three different values of $\beta$, representing two interesting regimes, and vary the coupling to characterize the effect of the coupling on the height of the gravitational wave spectrum.  In Fig.~\ref{fig_0.45amplitude_table} we compare the {\sl maximum} height of the gravitational wave spectrum $\Omega_{\rm GW, max}$ as a function of the coupling $\xi$, and show: (a) the total spectrum, (b) the spectrum sourced only by the fluid and (c) the spectrum sourced only by the field.  We repeat this procedure for $\beta = 0.01$, $\beta = 0.028$ and $\beta = 0.1$.  

For much of the parameter space we explored, the fluid contribution to the gravitational wave spectrum was unimportant -- the peak height was largely unaffected by the fluid source.  This is not to say that the fluid played no role, as the overall height of the spectrum in these cases was affected by the presence of the fluid, but this effect came from the way the fluid acted as a friction term in the field equations of motion and slowed the field down.

However, for large enough values of $\beta$ and $\xi$ we did see that the fluid source was important and even dominant.  Another important result here is the fact that, although the relative contributions to the gravitational wave spectrum differs for large values of the coupling, the overall height of the spectrum is not changed significantly relative to the uncoupled case.

\begin{figure}[htb]
\includegraphics[width=.45\textwidth]{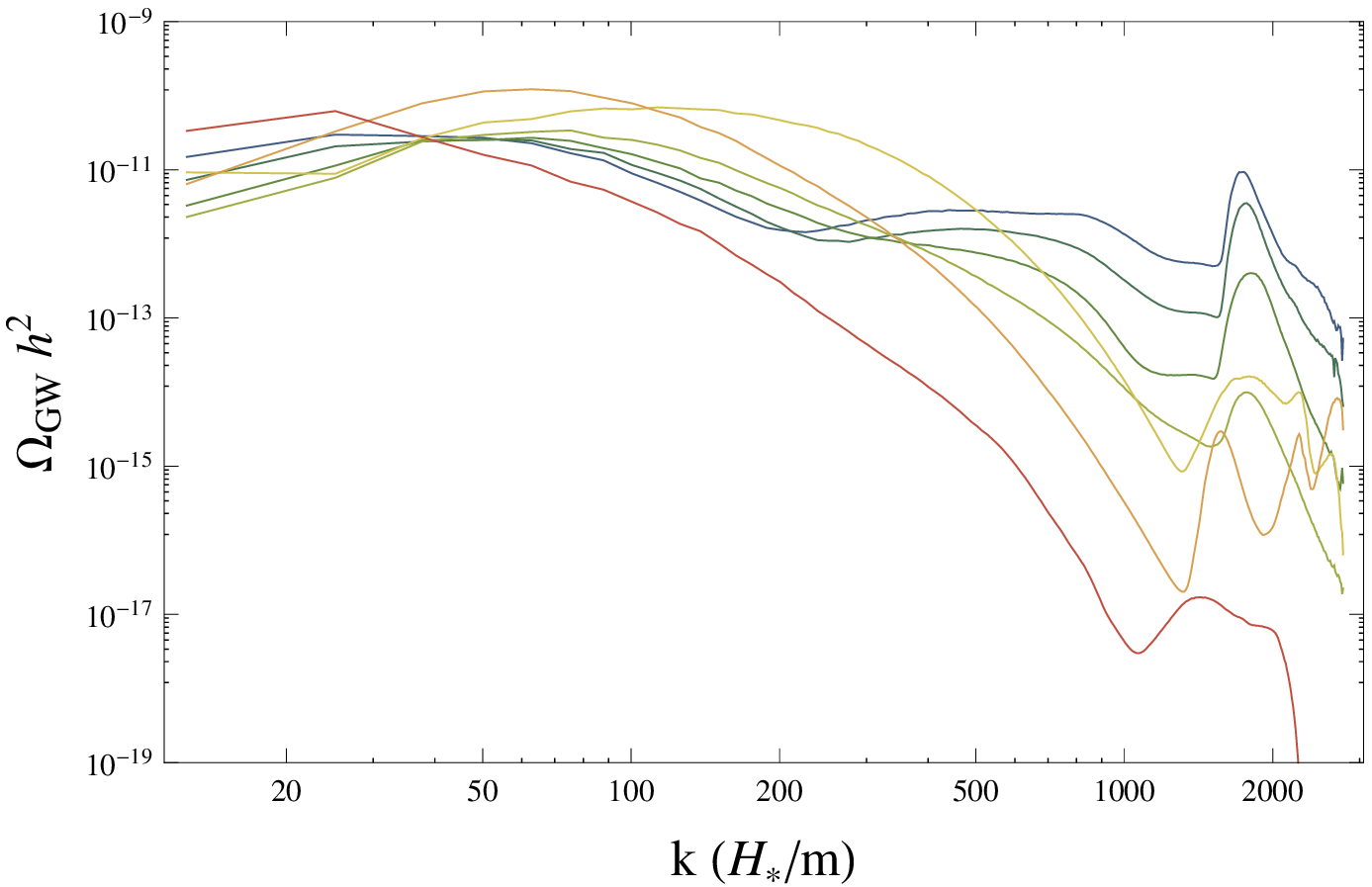}
\includegraphics[width=.45\textwidth]{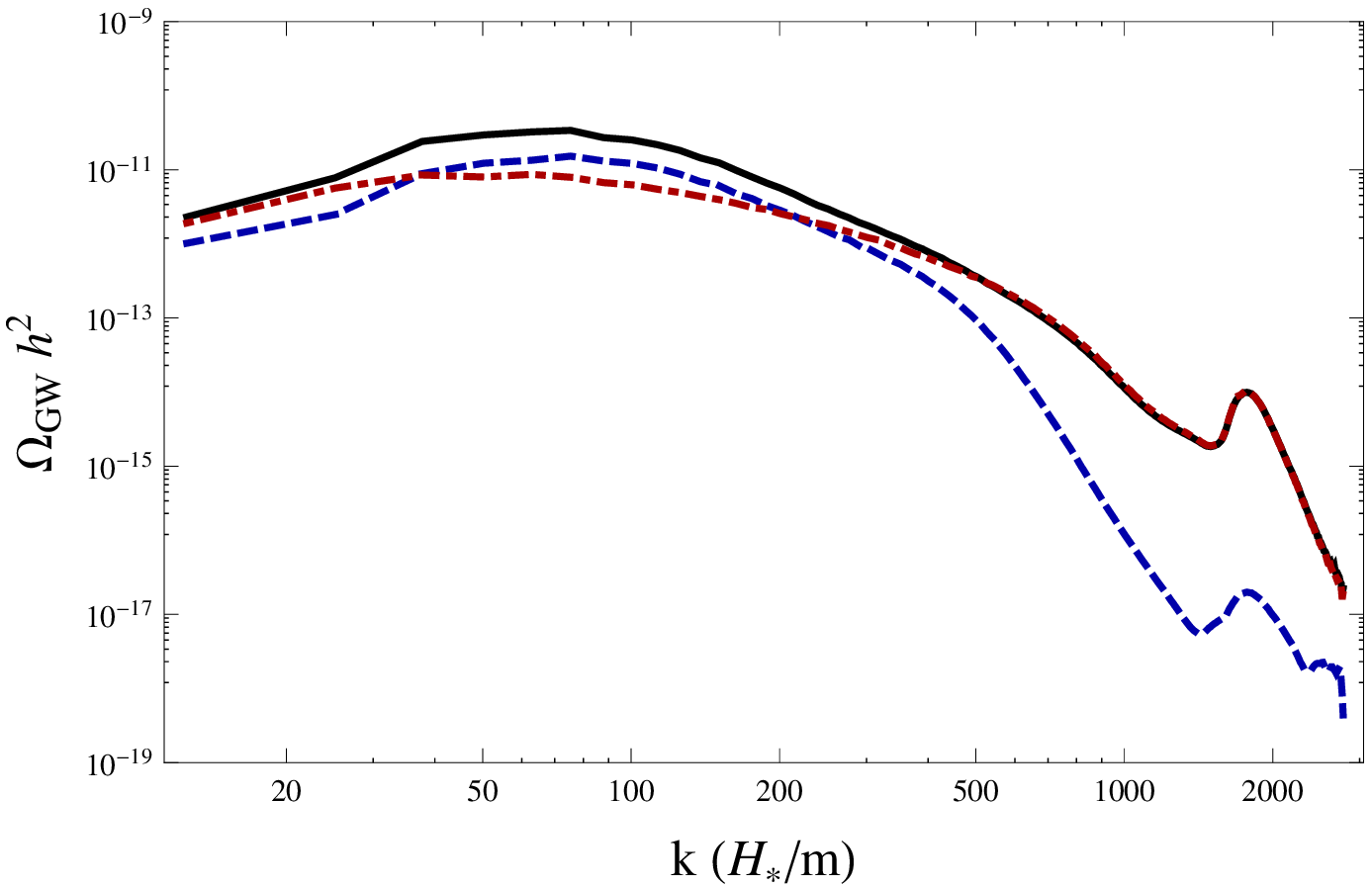}
\caption{\label{specfic45comp}The gravitational wave spectra for $\alpha = 0.45$, $\beta = 0.1$, and various couplings.  The upper panel shows the spectra at a time corresponding to $R(t) = 2.5 l_*$ and $\xi = 0.2, 0.4, 0.6, 0.8, 1.6, 3.2, 6.4$ from blue to red (roughly, top to bottom at $k~1000$).  The lower panel shows the contributions to the total gravitational wave spectrum (black, solid) from the fluid (red, dot-dashed) and the field (blue, dashed) for $\xi = 0.8$.}
\end{figure}

Lastly we would like to point out how the spectrum contains characteristic peaks from both the fluid source and the field source.  Fig.~\ref{specfic45comp} shows the spectra for a variety of couplings for a particularly interesting case, $\beta = 0.1$.  Here we show that there are actually three different scales that contribute to the final spectrum.  First, there's a scale associated with the collisions of the bubbles, estimated to be at $k \approx R(t)^{-1}$ where $R(t) = l_* \approx $ \cite{Kamionkowski:1993fg}.  This is the bubble collision peak and can be found at the lowest frequencies probed in our simulations,
\begin{equation}
k_{\rm bubble} \sim \frac{1}{l_*} \approx \frac{36R_0}{L}{6.28 R_0} \approx 4\pi \frac{H_*}{m}.
\end{equation}
To verify this, we do probe smaller frequencies (see Fig.~\ref{a.45res_comp}) to ensure that we correctly determine the amplitude at low frequency. 
There is additionally a scale from the coalescence of bubbles (a source that exists in the field sector) that should occur at somewhat higher frequencies \cite{Kamionkowski:1993fg,Child:2012qg}.  This should be on the order of the bubble wall thickness at the time of collision.  This effect appears for intermediate values, $k \sim 50 H/m$ of the frequency as in Fig.~\ref{specfic45comp}.  Lastly, we expect a contribution at high frequency, $k\sim 200 H/m$, due to the turbulence of the fluid.  This is a pronounced effect that becomes increasingly important at larger couplings.

In Fig.~\ref{specfic45comp}, a particularly interesting value of the coupling, $\xi = 0.8$, shows how all of these effects contribute to the final spectrum, at a time corresponding to $R(t) = 2.5 l_*$, or $t ~ 0.6 H_*^{-1}$ for this particular coupling.  This case was chosen as it had approximately equal-magnitude contributions from the fluid and field.

\subsection{The thin-wall, $\alpha=0.96$ case}
\label{96case}

As with the first case, we begin looking at the $\alpha = 0.96$ case by tying it to previous studies.  Here we take $\beta = 0.01$ throughout, as a maximum value of $\beta$ for which we have widely stable simulations. Even still, we increase our numerical resolution to $512^3$ for this section and use a volume $L = 10 R_0$ with $N_b = 16$ bubbles.

\begin{figure}[htb]
  \centering
    \includegraphics[width=0.45\textwidth]{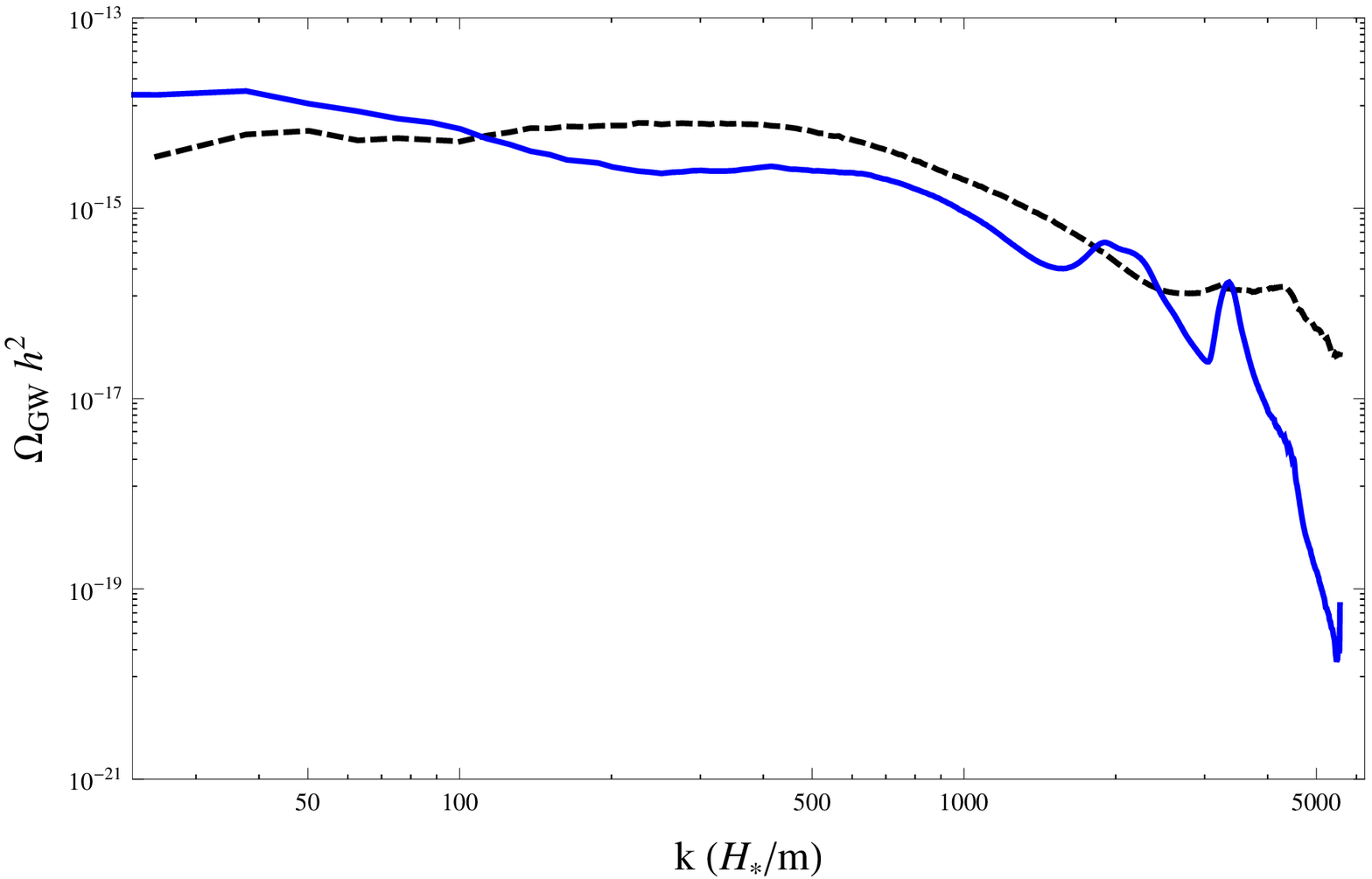}
  \caption{\label{C_G_comp} Comparison between our work with $\alpha = 0.96$, $\eta=0$ (solid, blue) and a similar simulation in \cite{Child:2012qg} (dashed, black).  These plots are evaluated at $t = 3 l_*$.  The overall amplitude is similar, but the shapes and peak frequencies disagree primarily due to the different values of $\alpha$ used (wall thicknesses differ by a factor of 2), and as the work in \cite{Child:2012qg} includes Hubble friction.  The amplitude from \cite{Child:2012qg} is scaled to match $\beta = 0.01$. }
\end{figure}

In Fig.~\ref{C_G_comp} we compare to the GUT-scale simulations probed in \cite{Child:2012qg}.  For this test we turn off the coupling ($\xi = 0$) while staying with $\alpha = 0.96$ (to be consistent with \cite{Giblin:2013kea}) rather than $0.98$ (the choice that is most similar to \cite{Child:2012qg}), and for a similar bubble spacing, $l_* \sim 2.2$.  We find approximate agreement between spectrum amplitudes, but somewhat different shapes and peak positions, presumably due to the small differences in bubble profiles and box sizes, and larger differences in well separation and Hubble friction.

\begin{figure}[tp]
\includegraphics[width=0.45\textwidth]{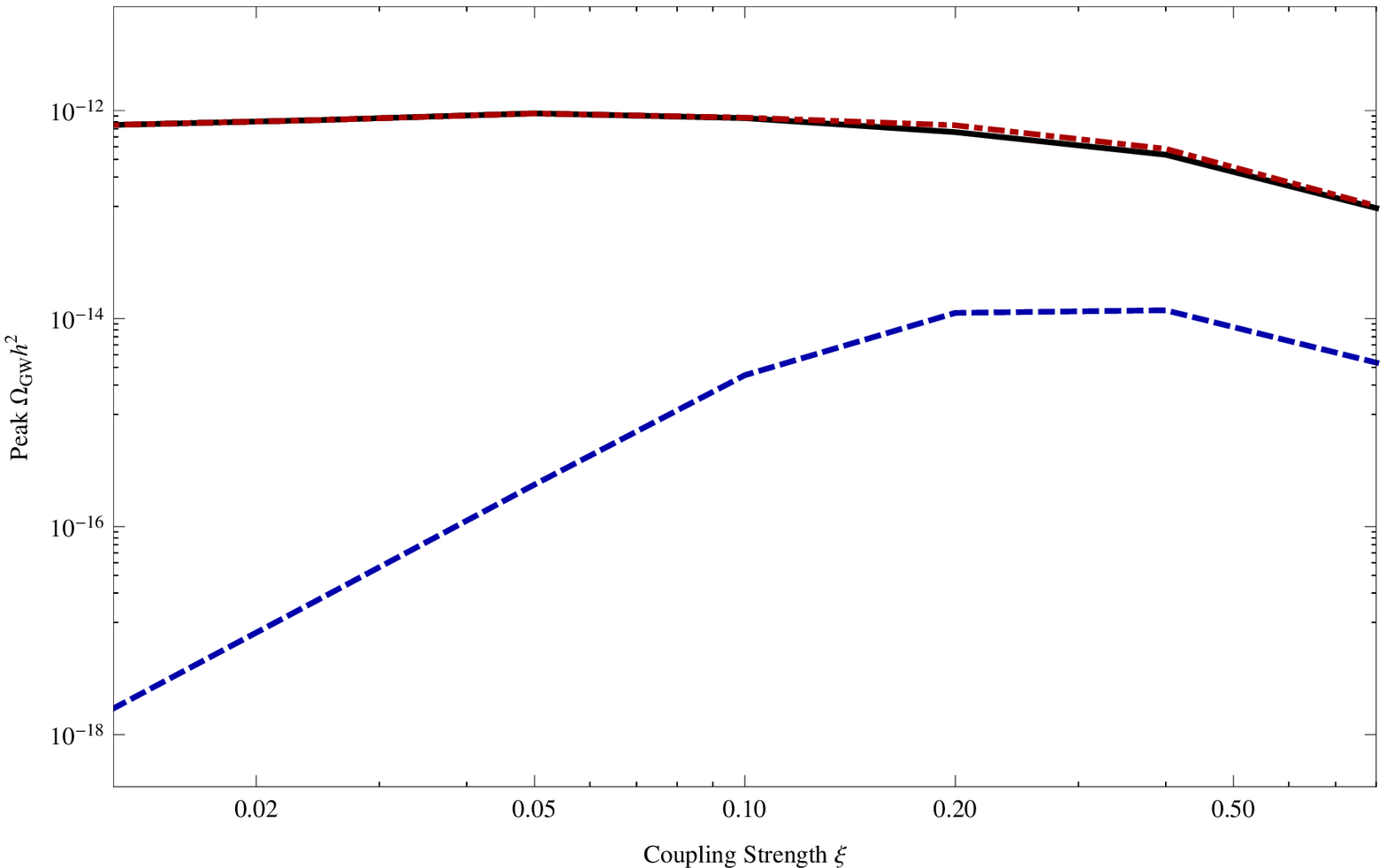}
\includegraphics[width=0.45\textwidth]{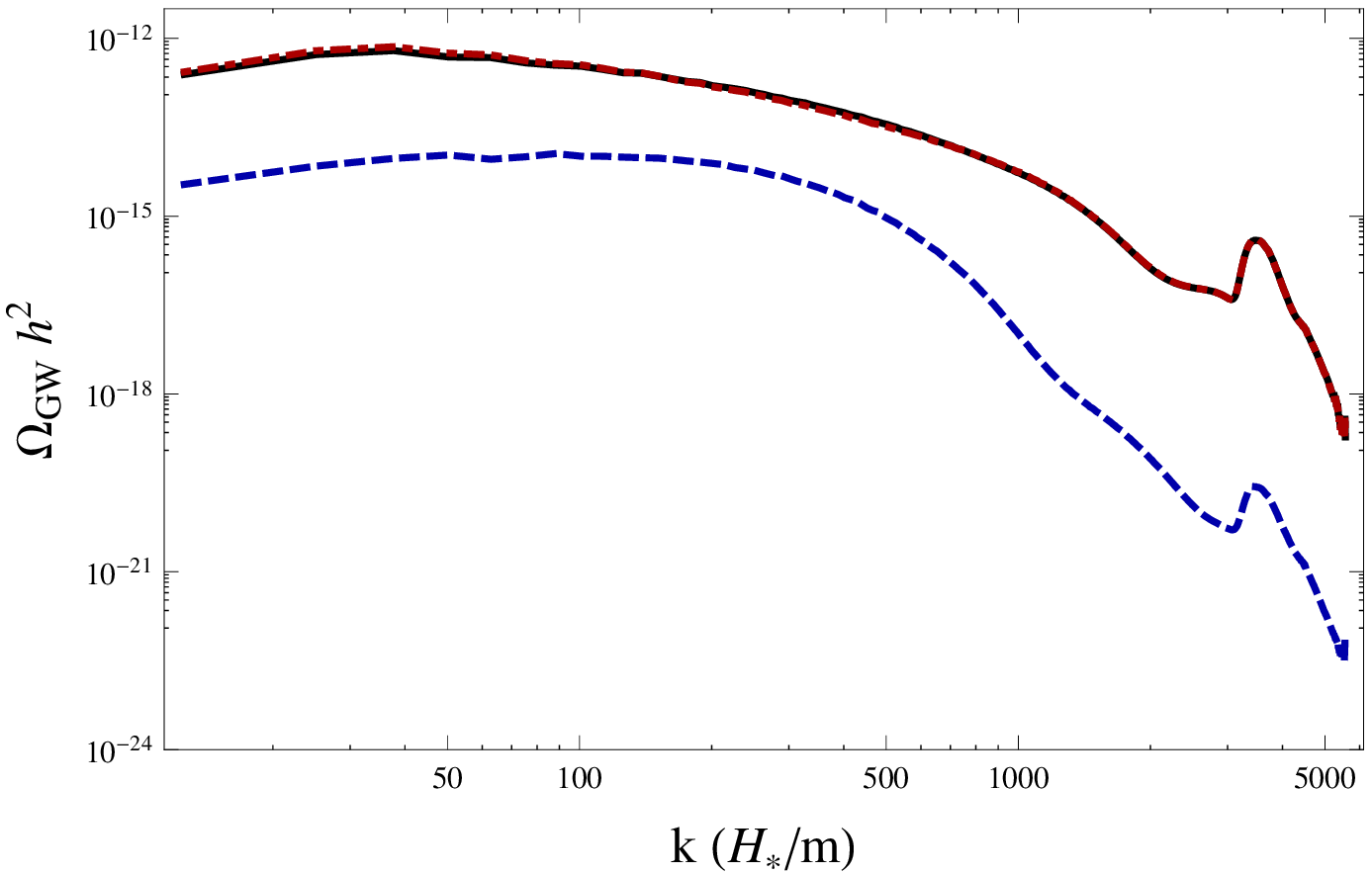}
  \caption{The upper plot shows amplitudes vs coupling for $\alpha = 0.96$ at $\beta = 0.01$ and $R = 1.5 l_*$, where all couplings examined are of the form $\xi = 0.0125 \times 2^n$.  The lower plot shows the gravitational wave specta for the $\xi = 0.2$ case.  The solid black line indicates the total power spectrum, red dot-dashed the field contribution, and blue dashed the fluid. \label{fig_0.96amplitude_table}}
\end{figure}

Next we look to probe whether the fluid can ever dominantly contribute to the gravitational wave spectrum for this type of phase transition.  In a similar manner to the $\alpha = 0.45$ case, we attempt to vary the coupling so that we see the effects for a wide range of possible phenomenology.  Fig.~\ref{fig_0.96amplitude_table} shows the relative contributions to the gravitational wave spectrum from the fluid and field for a particular choice of parameters.  Here we never see the fluid having a significant contribution to the energy density, although we still see that the height of the spectrum when sourced only by the field has a dependence on the coupling, even though the fluid does not dominate the spectrum.

To be clear, we are not saying that the contribution from the fluid is always sub-dominant; but we did not find it to be dominant for the specific coupling and range of parameters that we can probe.  Thin-wall simulations are much more difficult to keep stable, as the large field gradients tend to create numerical instabilities in the simulation.  Larger values of $\beta$ ($\beta > 0.01$) were also more difficult to explore for numerical reasons, a regime where we might expect larger or even dominant contributions from the fluid as in the $\alpha=0.45$ case.  

\section{Discussion}
\label{discussion}

In this work we showed that relativistic fluids and fields can both contribute to a background of stochastic gravitational waves sourced during a first-order phase transition.  We studied a range of phenomenological couplings between the field and fluid sector and showed that the production of gravitational radiation is robust.

We find that gravitational waves produced by bulk motions in the fluid can contribute appreciably, even dominantly, depending on the fluid-field coupling. As the coupling vanishes, the only contribution to the gravitational wave spectrum should be from the field, as little to no energy is converted into bulk motions of the fluid.  For larger couplings, more energy is deposited into bulk motions of the fluid, leading to an increasing contribution from the fluid sector.  Persistent perturbations in the fluid sector may continue to generate gravitational waves after coalescence (a result first noticed in \cite{Hindmarsh:2013xza} and can be seen explicitly in Figs.~\ref{fig_0.45amplitude_table} and \ref{fig_0.96amplitude_table}), a contribution which can even end up dominating the spectrum in some regions of parameter space.  

At the same time, in all cases that we probed, the most important impact on the gravitational waves produced was a modification of the spectrum sourced by the field.  This is somewhat surprising, since expectations often claim that the contribution from turbulence should often out-shine the contribution from the field; in our simulations we did not find this to be generically true, even in the presence of strong couplings.

We are mindful of our approximations, too: neglecting the expansion of the Universe, and by extension temperature-dependent effects, makes our simulations better approximations for fast phase transitions.  In the cases where the inter-bubble spacing is large and/or the couplings cause the bubbles to expand slowly, it is more important that these effects be taken into account, especially for larger couplings where the system may evolve slowly.

One can consider allowing the perturbations to persist for longer by decreasing the simulation box size $L$ relative to $H_*$, however this would scale the amplitude down as $(L/H_*)^2$, while we see the spectrum due to fluid perturbations grow only linearly, similar to \cite{Hindmarsh:2013xza}.  We therefore expect our amplitudes to correspond to near the maximum possible amplitude.

Many numerical estimates of current-day gravitational wave spectra see signals near, or around, $\Omega_{GW,0}h^{2} \lesssim 10^{-11}$, and we find ourselves solidly in that realm here.

\section{Acknowledgments}  We are particularly thankful to Glenn Starkman for useful conversations.  JTG is supported by the National Science Foundation, PHY-1068080.  JBM is supported in part by a Department of Education GAANN Fellowship and by the DOE Grant to the Particle Astrophysics Group at CWRU.

The majority of our simulations were conducted on hardware provided by the National Science Foundation, the Research Corporation for Scientific Advancement, and the Kenyon College Department of Physics.  Additionally, this work made use of the High Performance Computing Resource in the Core Facility for Advanced Research Computing at Case Western Reserve University and the Ohio Supercomputer Center under grant PKS0011-1.

\FloatBarrier

\end{document}